\documentclass[fleqn,usenatbib,useAMS]{mnras}

\usepackage{newtxtext,newtxmath}

\usepackage{graphicx}	
\usepackage{amsmath}	
\usepackage{amssymb}	
\usepackage{multicol}        

\usepackage{ae,aecompl}
\usepackage{graphicx}	

\def\beq{\begin{equation}}
\def\eeq{\end{equation}}

\def\arcmin{\mbox{$^\prime$}}%
\def\arcsec{\hbox{$^{\prime\prime}$}}
\def\fdeg{\hbox{$.\!\!^\circ$}}

\def\farcmin{\hbox{$.\mkern-4mu^\prime$}}
\def\farcsec{\hbox{$.\!\!^{\prime\prime}$}}
\def\sbmag{mag~arcsec$^{-2}$}

\title[A curved focal surface telescope for ULSB surveys]{Next-generation telescopes with curved focal surface for ultra-low surface brightness surveys}

\author[S. Lombardo et al.]{
Simona Lombardo,$^{1}$\thanks{E-mail: \texttt{simona.lombardo@lam.fr}}
Eduard Muslimov,$^{1}$
Gerard Lema\^{\i}tre,$^{1}$
and 
Emmanuel Hugot$^{1}$
\\
$^{1}$Aix Marseille Univ, CNRS, CNES, LAM, Marseille, France\\
}

\date{Accepted XXX. Received YYY; in original form ZZZ}
\pubyear{2019}

\begin{document}
\label{firstpage}
\pagerange{\pageref{firstpage}--\pageref{lastpage}}

\maketitle

\begin{abstract}
In spite of major advances in both ground- and space-based instrumentation, the ultra-low-surface brightness universe (ULSB) still remains a largely unexplored volume in observational parameter space. 
ULSB observations provide unique constraints on a wide variety of objects, from the Zodiacal light all the way to the optical cosmological background radiation, through dust cirri, mass loss shells in giant stars, LSB galaxies and the intracluster light. These surface brightness levels (>28-29 mag arcsec$^{-2}$) are observed by maximising the efficiency of the surveys and minimising or removing the systematics arising in the measurement of surface brightness. 
Based on full-system photon Monte Carlo simulations, we present here the performance of a ground-based telescope aimed at carrying out ULSB observations, with a curved focal surface design. Its off-axis optical design maximises the field of view while minimising the focal ratio. No lenses are used, as their multiple internal scatterings  
increase the wings of the point spread function (PSF), and the usual requirement of a flat focal plane is relaxed through the use of curved CCD detectors. 
The telescope has only one unavoidable single refractive surface, the cryostat window, and yet it delivers a PSF with ultra-compact wings, which allows the detection, for a given exposure time, of surface brightness levels nearly three orders of magnitude fainter than any other current telescope. 
\end{abstract}



\section{Introduction}
In the past decades the technological development has focused on improving the instruments and the techniques to test, e.g., the ${\Lambda}$CDM paradigm of galaxy formation and evolution on linear scales, such as the anisotropies of the Cosmic Microwave Background.
However the non-linear scales still show some discrepancies between theoretical predictions and current observations and remain to be tested.
One example is provided by the number of satellite galaxies \citep{bullock_2017} in the local group, which is order of magnitudes lower than what numerical simulations show.
One of the reason for such difference might lay in the ultra-low surface brightness (ULSB) of these dwarf galaxies (well below the ground-based sky
background), which would hence be missed by current surveys, not meant to observe very extended and faint structures.

Another discrepancy between prediction and observations can be found in the lack of an extended and irregular tidal tail around large galaxies which has been predicted \citep{Kazantzidis2008,cooper2010,cooper2013} but not significantly observed \citep{barton1997,fry_1999,Atkinson2013}.

Some progress has been made in the recent years and some of these features were observed \citep{delgado_2010,dokkum_2015,knapen_2017,mihos_2017,abraham_2014,duc_2015,boissier_2008}.
These surveys focus, however, on single objects or small regions in the sky, as they require long integration in order to reach the required signal-to-noise ratio (SNR), and a larger systematic study of the ultra-low surface brightness universe is still missing. 

In order to design the ideal survey for this type of observations, one would choose: a place with the lowest sky-background level; a telescope without obscuration, as any obscuration or spider causes diffraction and displace more power in the point spread function (PSF) towards the wings \citep{cabanac_2014,hugot_2014, muslimov_2017b}; a fast optical system with a small $f/D$ (where $f$ is the focal length and $D$ is the diameter of the entrance pupil), as for extended objects the imaging speed depends on the focal ratio (not on the aperture size, as for point-source objects); it should have either only refractive elements with anti-reflection coatings \citep[e.g.][]{abraham_2014}, or superpolished reflective surfaces, and no dust contamination.

A satellite with these characteristics would have the great advantage of drastically reducing the sky background and eliminating the molecular scattering due to the atmosphere (that degrades the PSF at large radial distances).
The introduction of filter coatings directly deposited on the CCD surfaces would additionally allow to observe in several bands including UV (observable only from space).
This is particularly important for observations of, e.g., the cosmic web filaments, believed to be brighter in UV wavelengths.
The cosmic web of filaments is supposed to contain the
missing fraction of baryons at low redshifts, therefore its observations would greatly improve our knowledge on the beginning and evolution of the universe.
This is for instance the basic concept behind the MESSIER space mission proposal \citep{valls_gabaud_2017}.
As one of the goal is observing in the UV, all refractive elements must be avoided (as they will generate Cherenkov emission by the abundant relativistic particles) and the design must be an unobscured fully reflective telescope with superpolished optics. 
 The proposed observation method is drift scan, which allows to achieve flat fielding accuracy of 0.0025\% \citep{Zaritsky_1996}.
 This  additionally requires that the instrument distortion \textit{has to be corrected at least in one direction}. 

Here we present a ground-based telescope (Section~\ref{sec:design}) that serves as a pathfinder and demonstrator to test and improve the curved sensor technology for astronomical usage.
The design of the pathfinder was previously shown by \cite{muslimov} and presents a primary mirror of 35.6 cm with a field of view of $>1.6^{o}\times2.6^o$. 
This telescope is a bi-folded fully reflective Schmidt with a curved focal surface.
The use of a curved CCD enhances the performances in terms of transmission and PSF shape, as it allows to simplify the overall system and (eliminates the need for field flattening lenses), while preserving the wide field of view \citep{rim_2008, blake_2013,guenter_2017}.
Here we perform a series of photon Monte Carlo simulations (Section~\ref{Sec:setup_Simul}-\ref{sec:simulations}) to verify its performances also at very large distances from the centre of the field of view.
We also simulate realistic observed fields and artificially inject some ULSB objects to test its full potential (Section~\ref{sec:field_simul}).

We are hence able to demonstrate that the pathfinder is a competitor to all current ULSB surveys and, as it will be mostly used in drift-scan mode, it will allow to observe a large fraction of the sky every night, as opposite to the other surveys that are focusing on single objects.

\section{The ground-based telescope demonstrator }

In the next Section we will briefly describe the design of the pathfinder.
Full scale simulations are required to asses the impact of possible scattered light and atmospheric effects over the full field of view of the telescope.
As our goal is the observation of objects that are even fainter than a typical sky-background, the evaluation of the PSF, particularly at the wings, becomes a main parameter to explore.
In this Section we also provide a description of the end-to-end photon Monte Carlo simulation software used.

\subsection{Telescope design}
\label{sec:design} 

\begin{figure}
 \begin{center}
  \includegraphics[width=0.45\textwidth]{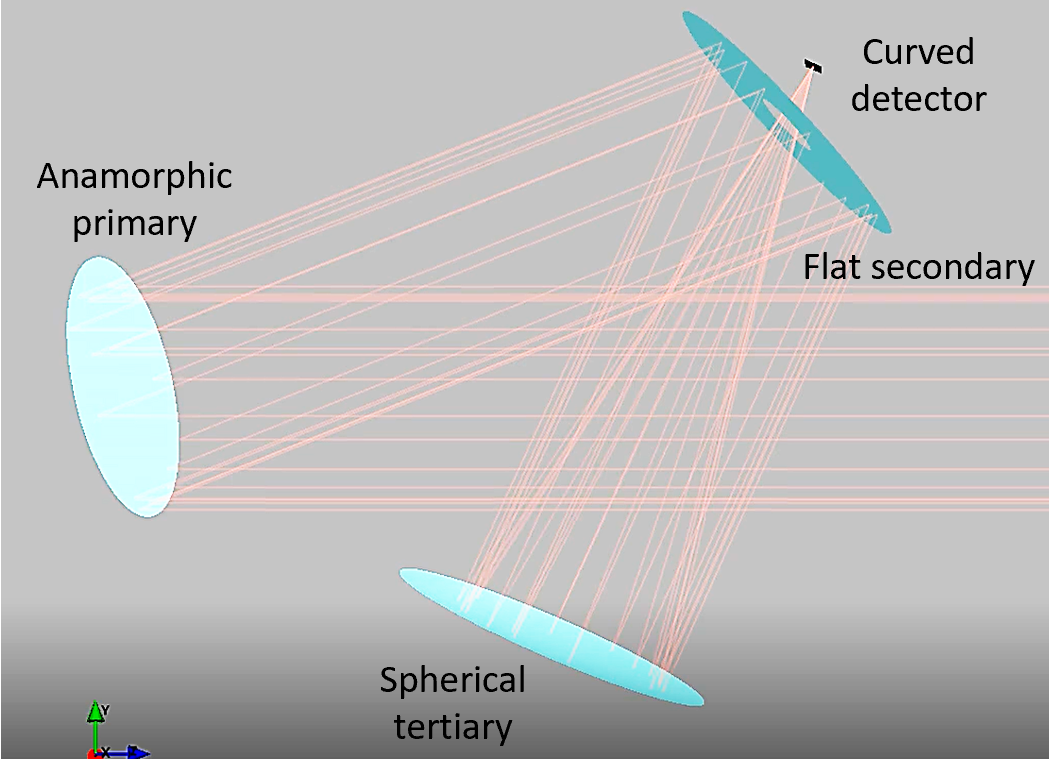}
  \caption{Optical design of the pathfinder: light reaches an anamorphic primary, is reflected by a flat secondary and
  a spherical tertiary yields a simple spherically-curved focal surface where a curved detector is placed.}
  \label{optical_design}
 \end{center}
\end{figure}

The ground-based pathfinder is a fully reflective Schmidt telescope \citep{muslimov}, composed of an anamorphic primary mirror of 35.6\,cm (whose purpose is to correct the spherical aberrations
), a flat secondary mirror and a spherical tertiary mirror that focuses the light onto a convex focal \textit{surface} (Figure~\ref{optical_design}).
The latter is equipped with a convex spherically-curved CCD with a radius of curvature of $\sim$900\,mm and 4030$\times$2480 pixels of 10~$\mathrm{\mu}$m pitch.

As we need to cool down and control the temperature of the CCD, the detector system is enclosed in a cryostat, whose window is the only, inevitable, refractive element of the system.
The bandpass filters (one or two large-band filters) are deposited directly on the window, to optimize the performances.
For simplicity we will only talk about a g-band filter with the same characteristics as the one planned for the Large Synoptic Survey Telescope \citep[LSST,][]{lsst_f_2008}, but more or different filters might be added in the future.

\subsection{Observing site}
\label{sec:site} %
Several dark sites are available in Europe  for the commissioning and installation of this demonstrator. La Palma and Calar Alto are the
best locations due to the quality of their skies \citep{leinert_1995,garcia_2010} and the accessibility of the observatory facilities. 
The choice of the site is particularly important to establish the main observing strategy, given the statistics of the atmospheric
parameters.
Both drift-scan and point-and-stare observing techniques are planned for the pathfinder, to allow for large sky coverage/per night (drift scan), or deeper observation of specific fields (point-and-stare) and
critically depend on the darkness of the site. In this paper we will
consider, as a proof of concept, a location within the 
Observatorio del Roque de los Muchachos in La Palma, for which there is a long record of atmospheric monitoring 
\footnote{\url{http://www.ing.iac.es/astronomy/observing/conditions/}}.

\subsection{End-to-end Monte Carlo simulation software}
\label{sec:endtotend}
The software used to perform the end-to-end photon Monte Carlo simulations is \texttt{PhoSim} \citep{peterson_2015} that allows to simulate photons from stars and galaxies, and raytrace them through the atmosphere, by creating seeing, atmospheric diffraction and extinction effects. 
Then the full optical system is realistically simulated including all the effects, such as misalignment, shifts, tilts and dust deposition on the optical surfaces.
\begin{figure*}
 \begin{center}
  \includegraphics[width=1.\textwidth]{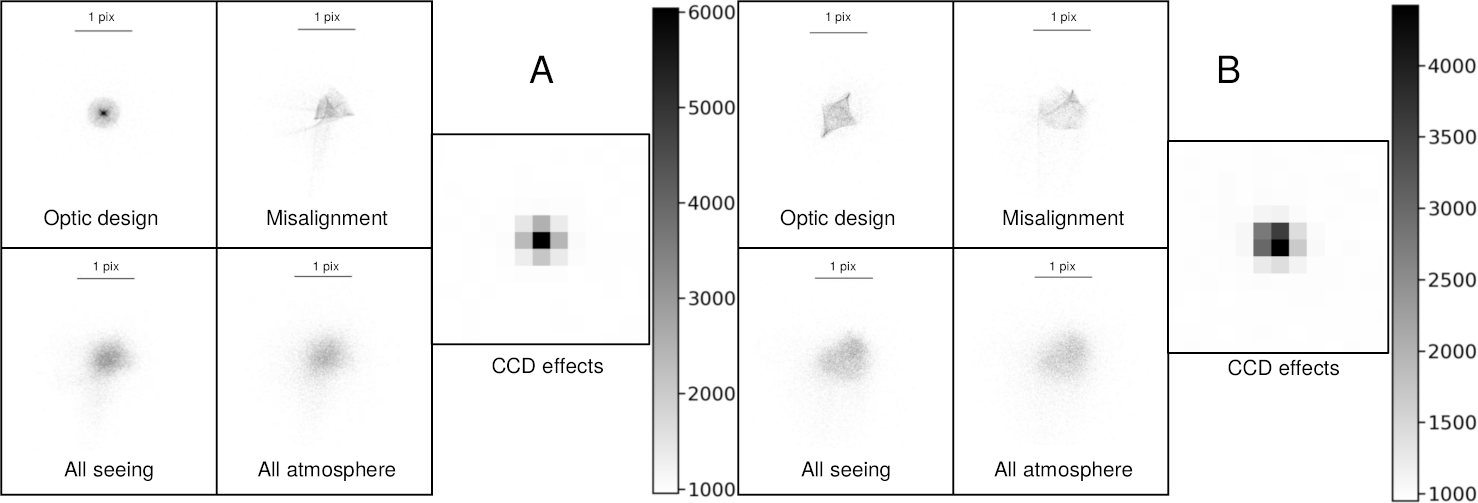}
  \caption{A: Photon Monte Carlo simulation of a star in the centre of the field of view, with \texttt{PhoSim}. The different simulation effects are added one after the other as follows. Upper left: optical design only. Upper right: perturbations, which include misalignment of surfaces and dust deposition. Lower left: seeing and dome seeing effects. Lower right: all atmospheric effects, including extinction, diffraction and clouds attenuation. Middle right: CCD image, including noise and gain conversion (values of counts are in ADU). B: same as before but with the star observed at the corner of the field of view, at a distance of 0\fdeg7$\times$1\fdeg2 from the centre.}
  \label{spot_diagram_center}
 \end{center}
\end{figure*}
Finally the detector is simulated with all the effects due to charges diffusion in the silicon layer of the sensor and the noise errors due to readout, dark current, pixel response non-uniformity, etc.

Figure~\ref{spot_diagram_center}A shows the spot diagram in the focal surface of the telescope of a star simulated in the centre of the field of view. 
All the previously mentioned effects are added one after the other, to illustrate the capabilities of the software and the quality of the optical design.
When simulating all the effects, most photons are still located within one pixel (10~$\mathrm{\mu}$m pitch, corresponding to 2.32" angular dimension on sky) of the detector at the focal surface. 
Note that the "star-shaped" image in the optical-design-only case in Figure~\ref{spot_diagram_center}A is due to the geometry of the distribution of rays in front of the entrance pupil of the telescope, and not to the presence of spiders holding the mirrors and obstructing the light.
These elements were carefully avoided in the design of the pathfinder as they bring more power to the wings of the PSF. The only obstruction is due to the hole in the secondary mirror.

The quality of the PSF is ensured also at the edge of the field of view, as shown in Figure~\ref{spot_diagram_center}B for a star observed at 0\fdeg7$\times$1\fdeg2 from the centre of the field of view. 
The image appears more deformed than in the previous case, but most rays are still focused within a single pixel.
The benefit of using a curved detector appears particularly evident at the edge of the field of view, where a better uniformity of the PSF is ensured with respect to the case where field flattening optics are used in combination with a flat detector \citep{muslimov}.

This is a key point to consider when the observing strategy includes the drift-scan. 
In this case the star is observed across the full field of view, and an highly distorted image at the corner of the field would result in more pronounced PSF wings.
The uniformity of the PSF on the full field of view is hence a parameter to consider and optimize already at the optical design stage.

The reliability of the photon propagation in the simulation software was tested by comparing the spot diagrams at the focal surface of the telescope in the optical-design-only case with the equivalent simulation in \texttt{ZEMAX}. 
These two provided the same results, as already verified for the LSST case \citep[][where it is also possible to find an accurate description of the physics involved in the simulation of the atmosphere]{peterson_2015}.

\section{Simulation setup and input parameters}
\label{Sec:setup_Simul}

\texttt{PhoSim} allows the user to modify the files used to set all the input parameters for the simulation. 
These files describe not only the optical design but also the observing site characteristics and location, which determine the atmospheric parameters.  
The input files describing typical atmospheric conditions have been prepared to accurately include the proposed observation site characteristics. 

The wind speed and direction data, used to simulate the atmospheric turbulence for La Palma, are estimated from \cite{ncep}.
In Table~\ref{tab: site_data} are indicated the location information and averaged values for some of the parameters. 
The seeing used is an intermediate value between the best (0\farcsec72) and the worst (1\farcsec3) seeing at La Palma observatories\footnote{\url{http://www.ing.iac.es/astronomy/observing/conditions/}}. 
The aerosol parameters are estimated from the averages of the year 2017 measured by AERONET aerosol robotic network\footnote{\url{http://aeronet.gsfc.nasa.gov/cgi-bin/draw\_map\_display\_aod\_v3}} for the Iza\~na location.

\begin{table}
\begin{center}
\caption{Site data and atmospheric parameters used for the simulations.}
\begin{tabular}{lr}
\hline\\[-1.5ex]
\multicolumn{2}{c}{Site and mean atmospheric data}\\
\hline\\
Latitude [W] &   28\fdeg76 \\[0.25ex] 
Longitude  &   17\fdeg89\\ [0.25ex]
Altitude  & 2326 m\\ [0.25ex]
Seeing  & 1\farcsec0 \\ [0.25ex]
Dome seeing  & 0\farcsec1\\ [0.25ex]
Aerosol optical depth & 0.061\\ [0.25ex]
Aerosol index & 0.85 \\ [0.25ex]
\hline\\[-1.1ex]
\end{tabular}
\label{tab: site_data}
\end{center}
\end{table}

Other parameters, e.g. dome seeing, for which there are no known data, have the same values as provided for the LSST example in the code. 
In \texttt{PhoSim} the surfaces of the optical elements presents some microroughness which scatters light incident at large angles and consequently distribute more power to the wings of the instrument PSF.
This feature is kept in the pathfinder simulations, so that we can study its performances and decide in a later stage if superpolished surfaces (where such microroughness effects will be strongly reduced) are necessary or can be avoided. 

To add more realism in the simulations, the telescope surfaces are misaligned through shifts and tilts whose probability distribution is given by \texttt{ZEMAX} tolerances file.
To reduce the number of refractive elements in the design, we coat the cryostat window with a $g$-band LSST-like filter, 
  at $\sim471.1$\,nm and with a bandwidth of $\sim123$\,nm.
 The window has a thickness of 5\,mm and it is made of silica.

A very important aspect to consider in the design of the pathfinder is the anti reflection (AR) coating to apply on the CCD.
The typical AR coatings available for CCDs have lower transmission than the AR coatings used for lenses and often they let >5\% of the light to be reflected back.
This means that for a classical Schmidt telescope design (on-axis and with the corrector plate) a large fraction of light is reflected back through the cryostat window and/or corrector plate and part of this will in turn be reflected in the CCD direction again, creating an unfocused image of the object observed.
These ghost images limit the observable highest magnitude and prevents to reach the required level for ultra-low surface brightness objects observation.

The pathfinder design is not immune to these features as the light reflected back from the CCD can pass trough the obscuration in the secondary mirror and be focused on the focal surface again by the spherical tertiary mirror.
However we can greatly limit this effect by choosing an appropriate AR coating for the CCD.
For the purpose of the simulation we selected the same AR coating used for the MUSE CCDs \citep{reiss_2012} which are e2v CCD231-84, and their quantum efficiency (QE) with reflectivity values $\sim$1.2-2.0\% around the central wavelength of the filter.  
The noise properties, such as the readout and dark current, have been selected using datasheet of CCD290-99 from Teledyne-E2V\footnote{\url{http://www.e2v.com/resources/account/download-datasheet/1897}} and are listed in Table~\ref{tab: ccd_data}.

\begin{table}
\begin{center}
\caption{CCD parameters used in the simulation.}
\begin{tabular}{lr}
\hline\\[-1.5ex]
\multicolumn{2}{c}{CCD properties}\\
\hline\\[-1.8ex]
Pixel pitch  &   10$\, {\mu}$m/2\farcsec32 on sky \\[0.25ex] 
Pixel number   & 4030$\, \times \,$2480  \\ [0.25ex]
Readout noise  & 2.5\,e$^-$ \\ [0.25ex]
Dark current  & 4\,e$^-$ pixel$^{-1}$ hour$^{-1}$  \\ [0.25ex]
Gain  & 1.7\,e$^-$ ADU$^{-1}$\\ [0.25ex]
Radius of curvature & 887.1\,mm  \\ [0.25ex]
Shape & convex  \\ [0.25ex]
\hline\\[-1.1ex]
\end{tabular}
\label{tab: ccd_data}
\end{center}
\end{table}

The impact of the curving process on these characteristics of the detectors has not been fully determined yet, while some manufacturer tested few prototypes and found increased values for the dark current \citep{gregory_2015}, others found no clear performance degradation with respect to the flat sensor case \citep{lombardo_2019}.

\section{Simulated Point Spread Function and observed field}
\label{sec:simulations}
The quality of the PSF of the telescope is a key aspect to consider in ultra-low surface brightness objects observations, as already discussed in previous publications \citep{abraham_2014}.
The telescope must provide PSF  whose wings are as low as possible, such that the faint emission of these objects is not dominated by the PSF residuals of brighter stars or galaxies in the same observed field.
\subsection{Point Spread Function from single star simulation}
\begin{figure}
 \begin{center}
  \includegraphics[width=0.50\textwidth]{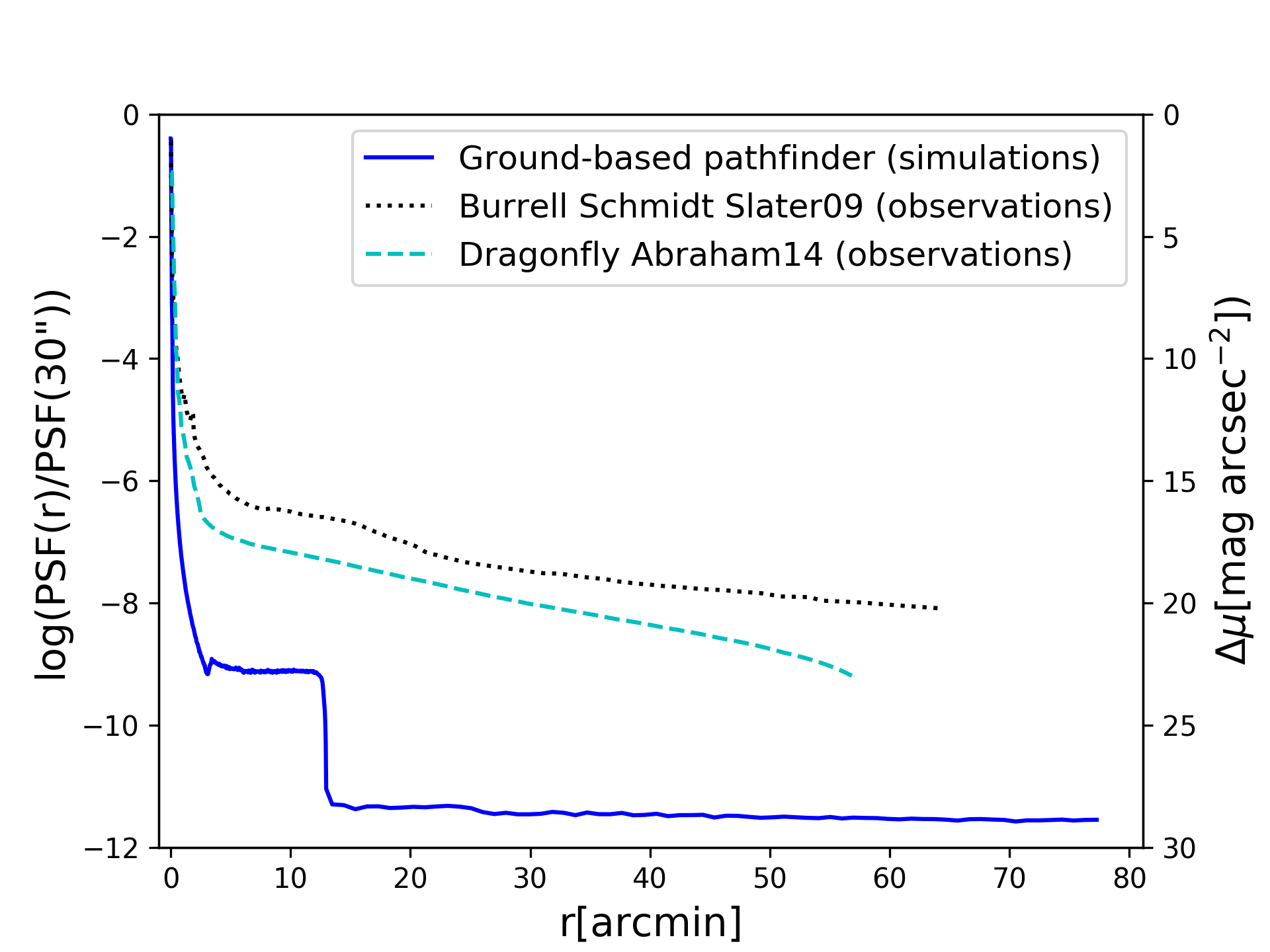}
  \caption{PSF of the ground-based pathfinder presented in this paper from photo Monte Carlo simulations in solid blue line, PSF of Dragonfly in cyan dashed line \citep{abraham_2014} and PSF of the Burrell Schmidt in dotted black line \citep{slater_2009}. All curves are normalized with their values within 30\arcsec. }
  \label{psf_center}
 \end{center}
\end{figure}

The PSF of the pathfinder is shown in Figure~\ref{psf_center} as function of the radial distance in arcmin.
This PSF is obtained from a simulation of a 9 mag star in the centre of the field of view and it includes all the effects of optical design perturbations, atmosphere and seeing (described in Section~\ref{Sec:setup_Simul}).
The PSF is computed from an image composed of 140 different observations where the CCD was exposed for 40.5\,s each time. 
For each of the images, we allowed the atmospheric parameters to variate randomly in the simulations.
Additionally, the simulation does not include any CCD effects as we do not want to be limited by any systematic effect for our PSF estimation (e.g., the saturation of the central pixel).

Each value of the PSF in Figure~\ref{psf_center} is an average of the number of photons at the focal surface that are inside the pixels located at the same distance from the centre of the star (for each ring). 
The two additional curves in Figure~\ref{psf_center} show the PSFs for Dragonfly and the Burrell Schmidt telescope \citep[respectively][]{abraham_2014, slater_2009}.
All curves are normalized within 30\arcsec  from the centre, and we can see how low the PSF wings are for the pathfinder.
This comparison can be used only as a qualitative estimation of the improvements achievable with a curved focal surface design optimized for ultra-low surface brightness observations, as the data for the pathfinder are purely extracted from simulations whereas the other two are estimates from real observations and hence include more systematic effects.
\begin{figure*}
 \begin{center}
  \includegraphics[width=0.90\textwidth]{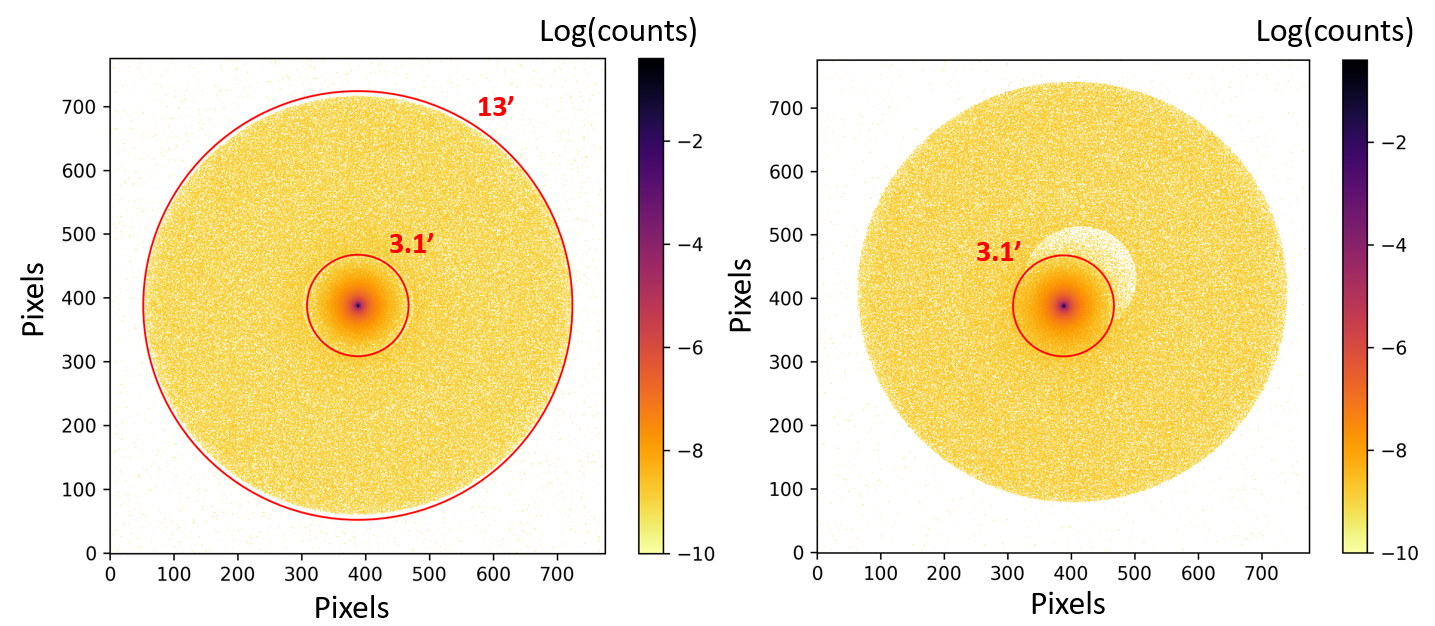}
  \caption{2D view of the PSF of the pathfinder shown in Figure~\ref{psf_center} for a star observed at the centre of the field of view (left) and at a distance of  0\fdeg7$\times$1\fdeg2 from the centre (right). The red circles represents radial distances of 3.\arcmin1 and 13\arcmin.}
  \label{psf_center_2d}
 \end{center}
\end{figure*}
\begin{figure}
 \begin{center}
  \includegraphics[width=0.52\textwidth]{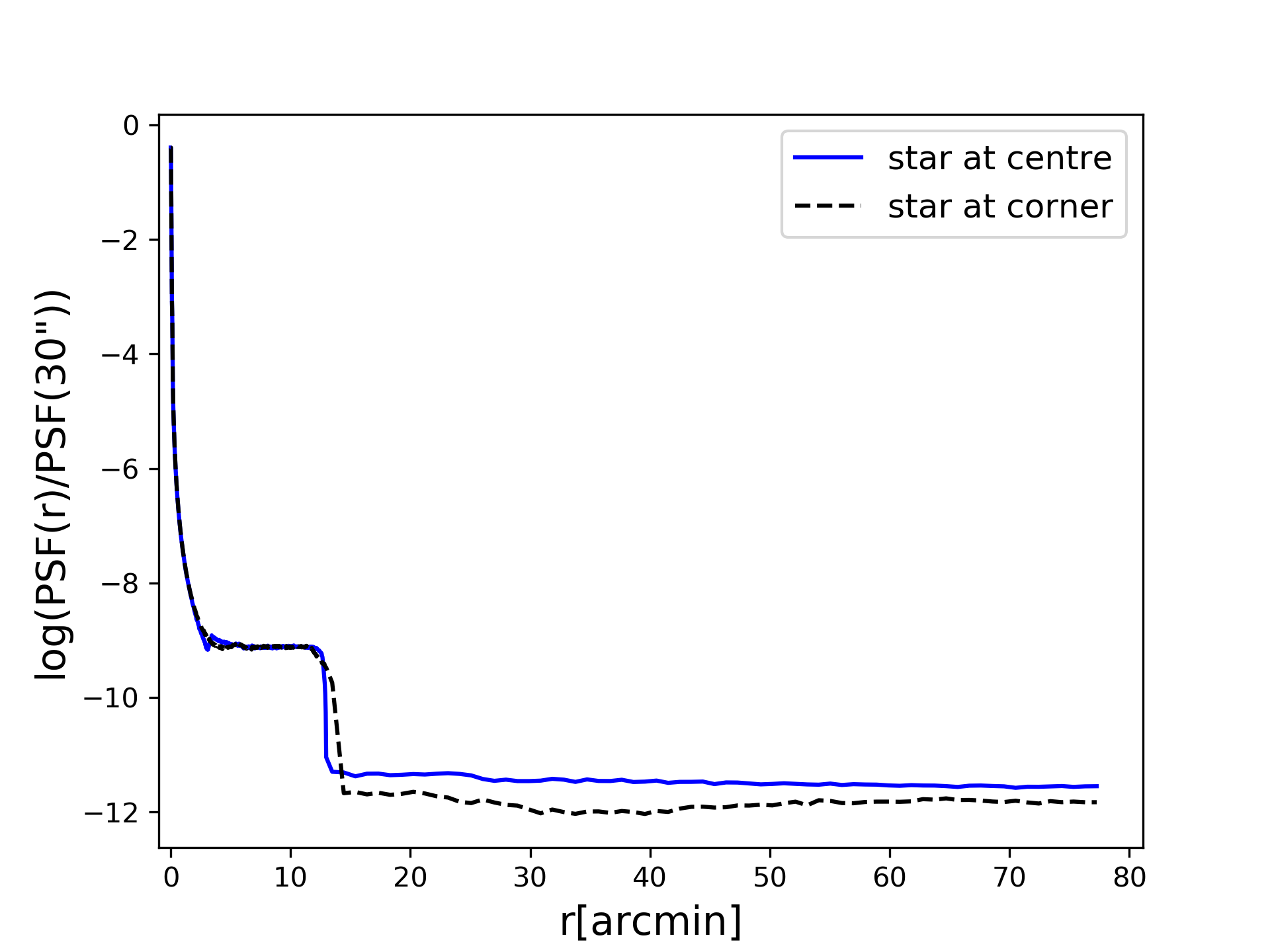}
  \caption{PSF of the ground-based pathfinder from photo Monte Carlo simulations of a star at the centre of the field of view of the pathfinder (blue full line) and at the corner  (0\fdeg7,1\fdeg2) (dashed black line). All curves are normalized with their values within 30\arcsec.}
  \label{psf_centershift}
 \end{center}
\end{figure}

The PSF of the pathfinder appears to decrease rapidly down to $10^{-9}$, where it plateaus. This plateau extends from a distance of 3' to 13' from the centre of the field of view, and it is caused by a ghost image of the star due to the back reflection from CCD and filter. 
The ghost, however, does not degrade the PSF as it is suppressed by 8 orders of magnitudes with respect to the centre.
After 13' the PSF decreases again reaching the $10^{-11.5}$ level.

A 2D view of the PSF is shown in Figure~\ref{psf_center_2d} (left image), where every square represents a pixel of the detector in the focal surface of the telescope.
From the Figure we can  distinguish the black central area, where most counts are, (67\% of the total number of photons are inside the central pixel and 98\% are within a circle of 5\farcsec8 radius) and the large  plateau that extends up to 13\arcmin.
Some of the photons simulated do not reach the focal surface at all as they first hit the primary mirror of the telescope with large angles and end up scattered outside immediately or after a few bounces on the other two mirrors of the telescope.

In Figure~\ref{psf_center_2d} we can also find the 2D image of the PSF for the star observed at 0\fdeg7$\times$1\fdeg2 from the centre of the field of view (right image). 
The PSF appears almost identical to the previous case (65\% of the total number of photons are inside the central pixel and 98\% are within a circle of 5\farcsec8 radius), with only the ghost image shifted with respect to the centre of the PSF. 
A better comparison between these two cases is shown in Figure~\ref{psf_centershift} where we can see that the two curves overlap. 
Since in Figure~\ref{psf_center_2d} (right image) and Figure~\ref{psf_centershift} the observed star is located at the corner of the field of view, where the highest amount of image deformation is expected, we can conclude that the PSF of the pathfinder is uniform across the full field of view.

\subsection{Simulation of a field in the sky}
\label{sec:field_simul}
The main science goal of the pathfinder is the observation of ultra-low surface brightness objects.
In order to test our capacity to observe such faint emissions, we simulate the entire telescope system in the case of an observational campaign of a sky field of 5\arcmin$\times$5\arcmin, this time also including the CCD effects. 

The field is composed of stars and galaxies drawn from the Millennium Simulation and generated using \texttt{CatSim}\footnote{\url{https://www.lsst.org/scientists/simulations/catsim}}.
In addition to this, a large elliptical galaxy (of 2\farcmin5 observed angular diameter) and an arch-like structure have been added to the simulation at the centre of the field.
The galaxy has an integrated brightness of 13.5 mag and the arches have surface brightness of 29 \sbmag~ which makes the full image similar to NGC5907 \citep{delgado_2010}.

In the simulations the field is observed at an airmass value of 1.13, for a total time of $\sim 30$ hours, which is equivalent to a week of observations in point-and-stare mode.
The observation is carried out through a series of short exposures of 382\,s ($\sim$6 min) and summed together to increase the contrast. 
This exposure time is equivalent to the passage of the star across the full sensor in a hypothetical drift-scan observation. 

\begin{figure}
 \begin{center}  \includegraphics[width=0.40\textwidth]{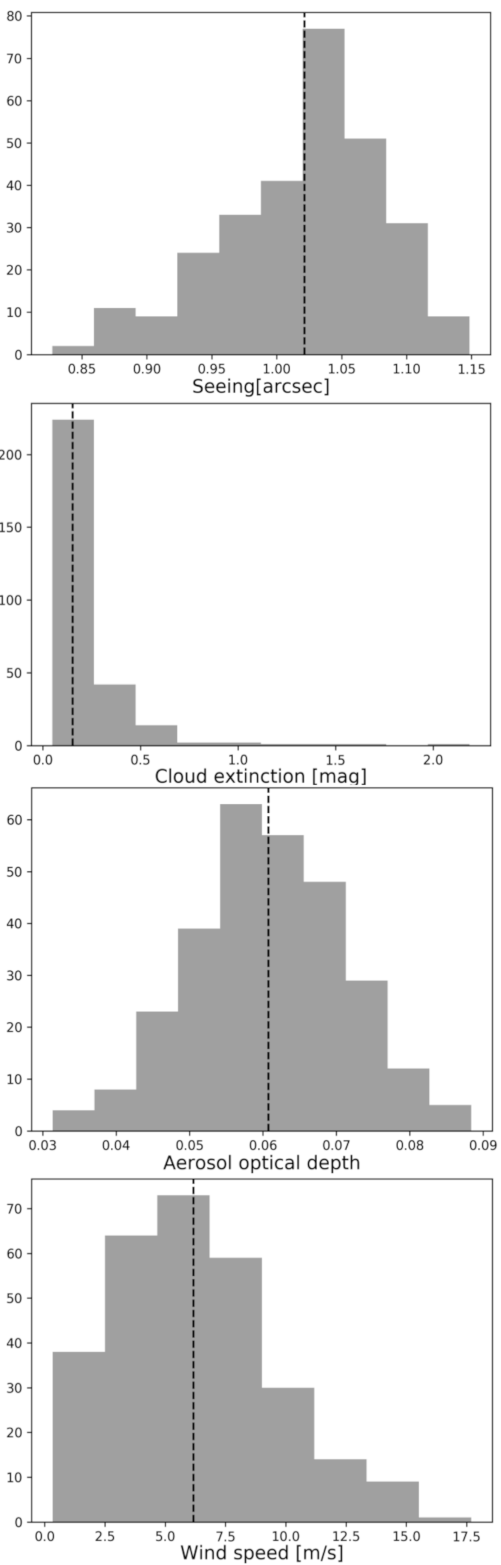}
  \caption{Histograms of the atmospheric parameter variations for the 288 shorter exposures that compose the 30 hours observation of the simulated 5\arcmin$\times$5\arcmin field. The vertical lines show the position of the mean value of the distributions.}
  \label{stat_field}
 \end{center}
\end{figure}

For each of these simulations a new set of atmospheric parameters and (this time also) CCD effects were used.
Figure~\ref{stat_field} shows the distribution of the parameters that mostly influence the atmospheric variation for the 288 exposures considered: for the seeing we have a mean value of 1\farcsec02$\pm$0\farcsec06, the median cloud extinction is 0.15\,mag, the average aerosol optical depth has a value of 0.061$\pm$0.010 and the wind speed is in average 6.2$\pm$3.3\,m s$^{-1}$. 

When we compare these values to the variations found within a year (excluding the months of June. July and August, which are the ones with the highest background emission in La Palma) at the  Observatorio  delRoque de los Muchachos in La Palma\footnote{\url{http://www.ing.iac.es/astronomy/observing/conditions/}}, we find that the simulated variations are smaller than the observed ones. 
The seeing shows standard deviations between 0\farcsec2 and 0\farcsec3, while the average value is around $\sim$0\farcsec93.
This means that the simulations are considering a more stable worst seeing condition with respect to what has been observed nearby the observatory.
Regarding the aerosol values measured at the Iza\~na location during the year 2017, they present variations of 0.2 against the 0.1 of the simulations. 
Even if there is a discrepancy, we can still say that the atmospheric parameters in the simulations well represent the variability of the observing conditions in a 9 months period. 

\begin{figure}
 \begin{center}  \includegraphics[width=0.40\textwidth]{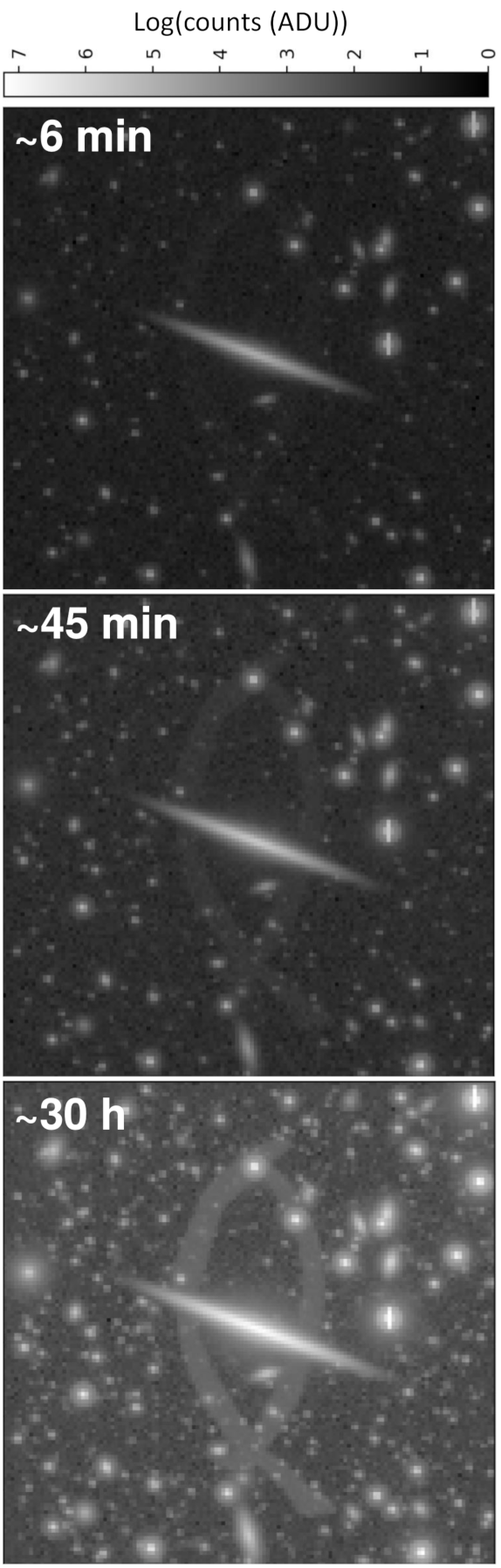}
  \caption{Simulation of a 5\arcmin$\times$5\arcmin field observed by the ground-based pathfinder after an exposure time of 6 min (top image), 45 min (middle image) and 30 h (bottom image).}
  \label{simulate_field}
 \end{center}
\end{figure}

The results of the simulations of the 5\arcmin$\times$5\arcmin  field observed at the centre of the field of view of the pathfinder are shown in Figure~\ref{simulate_field}.
At the top image a single frame of 6 min of integration is plotted, which corresponds to one night observation of drift-scan. 
Once the image is integrated for several consecutive nights (until it reaches 45 min of integration), the field appears more populated and better defined (middle image). 
It is possible to observe also the faint arch-like structure around the bright galaxy in the middle. 
This image is equivalent to a week of observations in drift-scan mode.
 
If one continues to integrate until the total exposure time is 30 h, corresponding to $\sim$10 months of drift-scan or 1 week of point-and-star observation mode, more objects appear and the arch-like structure is more visible.
In the drift-scan case the $\sim$10 months represent the total number of days required for the observations, and it will be spread over a few years depending on the specific object visibility in the sky. 

The signal to noise ratios (S/N) obtained per pixel in the ULSB object are $\sim$29, $\sim$7, and $\sim$4 for the 30 h, 45 min, and 6 min integration time images, respectively.
These are computed by rerunning the simulations with only the ULSB object in the field with the same atmospheric conditions used for the full field simulations. 
The median signal across the pixels composing this ULSB object is, then, extracted and divided by the square root of the median signal of the ULSB object from the full field image (which contains not only the faint object but also the noise and the background noise due to the presence of background stars and galaxies). 
The pixels of the ULSB object overlapping with the bright stars in the full field simulations are excluded from both calculations. 
These results show that we are able to observe, in a fraction of sky equivalent to 1 pixel (5.4 arcsec$^2$), a feature with surface brightness of 29.00 \sbmag~ at a precision of $\sim$3\%, $\sim$13\% or $\sim$26\% depending on the exposure time used.

The presence of the ULSB object in the images implies the capability of the pathfinder of observing such extremely faint and extended objects and of reaching a good S/N, even after just one week of nightly integration time. 

\section{Conclusions}

The ultra-low surface brightness universe still remains a niche for observations, as a full-sky survey is missing.
As observations  of this kind demand a telescope with an highly optimized design, we propose here an alternative concept that uses a curved detector. 
The advantage of such telescope is due to its unique design that combines the wide field of view of a Schmidt telescope, with the compactness, the lack of supporting spiders and field flattening optics provided by the usage of a curved CCD to match its curved focal surface. 
All these elements contribute to have a Point Spread Function with very low wings, key element for the observation of the ultra-faint universe.

The proposed telescope is a fully reflective, off-axis Schmidt design, with an anamorphic primary mirror of 35.6\,cm diameter, a flat secondary mirror and a spherical tertiary mirror that focuses the light onto a convex spherically-curved CCD.
In this paper we tested the full system through full scale photon Monte Carlo simulations, with the software \texttt{PhoSim}.
This software allows to simulate all the effects due to Earth atmosphere, the effects of misalignment of the optics and the propagation of the photon in the CCD, including the noise effect of the detector itself. 

These full scale simulations are the only way to study the illumination distribution of the full focal surface of the telescope and are mandatory to ascertain the potentiality of the pathfinder and the capability to reach the extremely low brightness level in the observations.
The results from the simulation of a star at the centre and at the corner of the field of view have shown firstly that the PSF is uniform across the full field of view, and secondly that the wings of the PSF reach unprecedentedly low level.
The normalised PSF decreases down to $10^{-11.5}$ at a radial distance from its centre of $\sim$13\arcmin which correspond to a difference in mag arcsec$^{-2}$ of 27 with respect to its value at the centre. 
Before reaching these values, the PSF shows a plateau of $\sim10^{-9}$  due to the ghost image of the star itself. 
The light of the star is in fact reflected back from the CCD and the filter and than it is propagated again in the direction of the focal surface reaching the CCD surface in a displaced position. 
This unfocused image of the star creates an halo of photons that extends from $\sim$3\arcmin  to $\sim$13\arcmin  from the centre of the PSF.
The PSF quality is not degraded by the presence of such ghost, as it is suppressed by 8 orders of magnitude with respect to the PSF central value.

Finally we simulated a field of galaxies and stars of 5\arcmin$\times$5\arcmin observed at the centre of the field of view for a total exposure time of 30 hours.  
This exposure time is reached by adding up 288 shorter observations of 382\,s ($\sim$6 min) and for each of these sub-exposures the atmospheric conditions have been changed randomly. 
The final image clearly shows the presence of the extended structure injected in the simulation and that is faintly emitting at 29 mag/arcsec$^2$, typical brightness for the ultra-low surface brightness objects. 
These results illustrate the full potentiality of the pathfinder, that will not only be used to test the groundbreaking technology of the curved sensors, but also provide important science outcome.

The design concept shown in this paper can, hence, serve as alternative to all current surveys for ULSB observations, or more in general targeted to extended objects. 
The installation of more telescopes of this kind could in principle extend the current field of view and photon collection capability and could provide a better sampling of the sky, especially if located also in the Southern hemisphere.

\section*{Acknowledgements}

The authors acknowledge the support of the European Research council through the H2020 - ERC-STG-2015 -- 678777 ICARUS program, the MERAC foundation and the European Astronomical Society who are both funding the project. 
The authors would like to thank David Valls-Gabaud and Francisco Prada, for their contribution to the project through fruitful discussions, and the referee for the constructive comments that improved the manuscript. This activity was partially funded by the French Research Agency (ANR) through the LabEx FOCUS ANR-11-LABX-0013. 




\bibliographystyle{mnras}
\bibliography{reference} 





\bsp	
\label{lastpage}
\end{document}